\documentclass{PoS}

\pdfoutput=1

\usepackage[utf8]{inputenc}
\usepackage{amsmath}
\usepackage{upgreek}
\usepackage{graphicx}
\usepackage[export]{adjustbox}

\newcommand\nd{n_{\rm domain}}
\newcommand\Nd{N_{\rm domain}}
\newcommand\Nc{N_{\rm core}}

\DeclareMathOperator\ceil{ceil}

\title{QPACE 2 and Domain Decomposition on the Intel Xeon
  Phi\thanks{Work supported by DFG in the framework of SFB/TRR-55.}}

\ShortTitle{QPACE 2 and Domain Decomposition on the Intel Xeon Phi}

\author{Paul Arts$^a$, Jacques Bloch$^b$, Peter Georg$^b$, Benjamin
  Gl\"assle$^b$, Simon Heybrock$^b$, Yu~Komatsubara$^c$, Robert
  Lohmayer$^b$, Simon Mages$^b$, Bernhard Mendl$^b$, \mbox{Nils
    Meyer$^b$,} Alessio Parcianello$^a$, Dirk Pleiter$^{b,d}$, Florian
  Rappl$^b$, Mauro Rossi$^a$, Stefan Solbrig$^b$, Giampietro
  Tecchiolli$^a$, \speaker{Tilo Wettig}$^{\,b}$, Gianpaolo Zanier$^a$\\
  $^a$Eurotech HPC, Via F. Solari 3/A, 33020 Amaro, Italy\\
  $^b$Department of Physics, University of Regensburg, 93040
  Regensburg, Germany\\
  $^c$Advanet Inc., 616-4 Tanaka, Kita-ku, Okayama 700-0951, Japan\\
  $^d$JSC, J\"ulich Research Centre, 52425 J\"ulich, Germany\\
  E-mail: \email{tilo.wettig@ur.de}}

\abstract{We give an overview of QPACE 2, which is a custom-designed
  supercomputer based on Intel Xeon Phi processors, developed in a
  collaboration of Regensburg University and Eurotech.  We give some
  general recommendations for how to write high-performance code for
  the Xeon Phi and then discuss our implementation of a
  domain-decomposition-based solver and present a number of
  benchmarks.}

\FullConference{The 32nd International Symposium on Lattice Field Theory,\\
  23-28 June, 2014\\
  Columbia University, New York, NY}

\begin{document}

\section{Introduction}

After the invention of lattice QCD in 1974 \cite{Wilson:1974sk} it
quickly became clear that (i) it had great potential to yield
physically relevant nonperturbative results and (ii) enormous
computational resources would be required to obtain them.  Such
resources were not readily available at the time.  This led some of
the physicists involved to design and build special supercomputers
optimized for lattice QCD simulations. These activities started in the
1980s and are continuing until now.  The resulting machines have been
serving as work horses for large-scale lattice QCD simulations
worldwide and were typically much more cost-efficient than
commercially available machines. They also influenced the design of
commercial supercomputers such as BlueGene. A probably incomplete list
of such machines includes ACPMAPS, GF11, Fermi-256, QCDSP, QCDOC (all
in the US), QCDPAX, CP-PACS, PACS-CS (Japan), as well as APE100,
APEmille, apeNEXT, and QPACE 1 (Europe), see also \cite{Wettig:2005zc}
for a review.

However, in the past several years the situation has changed. First,
the development of specialized processor chips has become too
expensive for academic projects. Second, commercial supercomputers
such as BlueGene scale well for lattice QCD and are reasonably cost-
and energy-effective (although the development of truly scalable
machines such as BlueGene or the K computer relies on enormous
government funding that is not always forthcoming).  Finally, standard
compute clusters (typically with accelerators) are now readily
available and can be used for capacity-type problems where strong
scalability is not a must.

Nevertheless, it still makes sense to pursue academic
hardware-development activities by combining commercially available
components in an innovative way, in particular regarding the
communication between processors.  This can result in machines that
are more scalable and more cost- and energy-effective than commercial
machines.  The development activities typically involve industry
partners that also benefit from the innovative concepts originating on
the academic side (``co-design'').  In our most recent projects these
partners were IBM (QPACE 1 \cite{Baier:2009yq}) and Intel (QPACE 2) as
well as Eurotech (QPACE 1 and 2).  Note that the current ``custom''
designs for lattice QCD are suitable for a broader application
portfolio than they used to be in the past.  Note also that hardware
development is only part of the story and that system software and
high-performance application codes are equally essential (and in fact
absorb a large fraction of the development activities).

In Sec.~\ref{sec:qpace2} we give an overview of the QPACE 2 project,
which is based on the Knights Corner (KNC) version of the Intel Xeon
Phi processor.  We also discuss in some detail our experiences with
this new ``many-core'' architecture and give recommendations for
programming it.  In Sec.~\ref{sec:DD} we present a new solver code
tailored to the KNC.  This code is based on domain decomposition (DD),
which is less bandwidth bound and more latency tolerant than other
approaches, and thus leads to better scaling behavior
\cite{Luscher:2003qa}.  In Sec.~\ref{sec:concl} we summarize and give
an outlook to QPACE 3, which will be an upgrade of the QPACE 2
prototype system, based on the future Knights Landing (KNL) version of
the Xeon Phi.

\section{Overview of QPACE 2}
\label{sec:qpace2}

Between QPACE 1 and QPACE 2, the acronym QPACE changed its meaning
from ``QCD Parallel Computing on the Cell Processor'' to ``QCD
Parallel Computing Engine''.  The QPACE 2 project is funded by the
German Research Foundation (DFG) and led by the University of
Regensburg in collaboration with Eurotech (Italy/Japan) and Intel,
with additional contributions from the J\"ulich Supercomputing Center
and the University of Wuppertal.\footnote{The development of QPACE 2
  was initially pursued in collaboration with T-Platforms.
  Unfortunately, this collaboration was terminated after T-Platforms was
  placed on the so-called Entity List of the US Department of Commerce
  in March 2013 and no longer had access to US technology, such as
  Intel Xeon Phi processors.  This caused a significant delay for
  QPACE 2.  Note that T-Platforms was removed from the Entity List in
  December 2013.}

\subsection{Node design}

\begin{figure}
  \centering
  \includegraphics[valign=c]{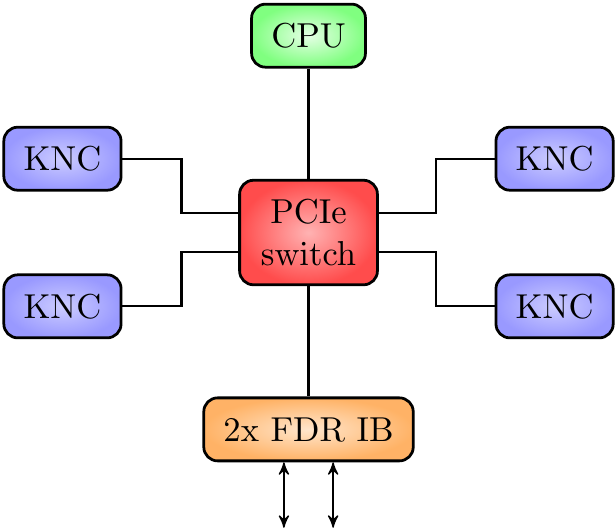}\hfill
  \includegraphics[valign=c]{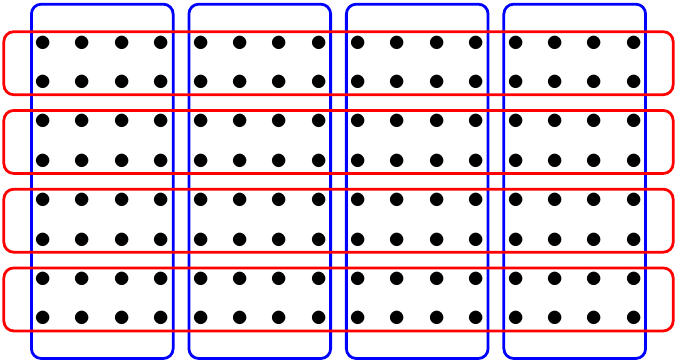}
  \caption{Left: QPACE 2 node design, see text for details.  Right:
    Sketch of the two-dimensional hyper-crossbar network (based on FDR
    Infiniband) of QPACE 2, here shown for $16\times8$ nodes as an
    example.  Each dot represents a node with two IB ports that is
    connected to one IB switch each in the $x$- and $y$-directions.
    The switches, which here have 32 ports each, are indicated by the
    red and blue squares, respectively.}
  \label{fig:node}
\end{figure}

Like most supercomputers, QPACE 2 consists of many identical nodes
connected by a network.  The node design is shown in
Fig.~\ref{fig:node} (left).
Four Intel Xeon Phi 7120X processors (a.k.a.\ Knights Corner or KNC)
are connected to a PCIe switch (PEX 8796 by PLX).  Also connected to
the switch are a low-power CPU (Intel Xeon E3-1230L v3) and a
dual-port FDR Infiniband card (Mellanox Connect-IB).  The CPU serves
as the PCIe root complex.  The KNCs as well as the IB card are PCIe
endpoints.  Peer-to-peer (P2P) communication between any pair of
endpoints can take place via the switch.  The rationale behind this
node design is that a high-performance network is typically quite
expensive.  A ``fat'' node with several processing elements and cheap
internal communications (here over PCIe) has a smaller
surface-to-volume ratio and thus requires less network bandwidth per
floating-point performance, which lowers the relative cost of the
network.  The number of KNCs and IB cards on the PCIe switch is
determined by the number of lanes supported by commercially available
switches and by the communication requirements within and outside of
the node.  We are using the largest available switch, which supports
96 lanes PCIe Gen3.  Each of the KNCs has a 16-lane Gen2 interface
(corresponding to a bandwidth of 8 GB/s), while both the CPU and the
IB card have a 16-lane Gen3 interface (i.e., almost 16 GB/s each).
The external IB bandwidth for two FDR ports is 13.6 GB/s.  This
balance of internal and external bandwidth is consistent with the
communication requirements of Lattice QCD, see also Sec.~\ref{sec:DD}.
Each of the KNCs has 61 cores, a clock speed of 1.238 GHz, a peak
performance of 1.2 TFlop/s (double precision), and 16 GB of GDDR5
memory with a peak bandwidth of 352 GB/s.  All of these numbers are
nominal.  Sustained numbers will be discussed in Sec.~\ref{sec:micro}.

One advantage of our node design is that the KNCs could in principle
be replaced by GPUs, although in this case a stronger CPU would be
advisable.

\subsection{Network}

The nodes are connected by an IB-based network with a hyper-crossbar
topology.  This topology was introduced by the CP-PACS collaboration
\cite{Iwasaki:1993me}.  For QPACE 2 we use a two-dimensional version
as indicated in Fig.~\ref{fig:node} (right).  In general, if we have
nodes with $d$ ports each and switches with $p$ ports each, a
$d$-dimensional hyper-crossbar has a maximum size of $p^d$ nodes.  We
are using IB edge switches with 36 ports, $p=32$ of which are used for
connecting the nodes.  Thus our maximum partition size is 1024 nodes,
corresponding to 4096 KNCs (which is sufficient for present-day
lattice sizes and more than our budget can afford).  The remaining 4
switch ports can be used for connecting the machine to a storage
system and/or for connecting the switches in a torus (using the
torus-2QoS routing engine in the OpenSM subnet manager).

The advantage of a hyper-crossbar over a torus is that we have full
connectivity in every single dimension with one switch hop, and
all-to-all connectivity by going through at most one intermediate
node.  The advantage over a (fat) tree is lower cost.  We note for
completeness that a higher-dimensional application can be mapped to a
lower-dimensional hyper-crossbar in a variety of ways.  Typically this
should be done such that the communication requirements are minimized.

\subsection{System design}

\begin{figure}
  \centering
  \raisebox{0mm}{\includegraphics[height=25mm,rotate=40]{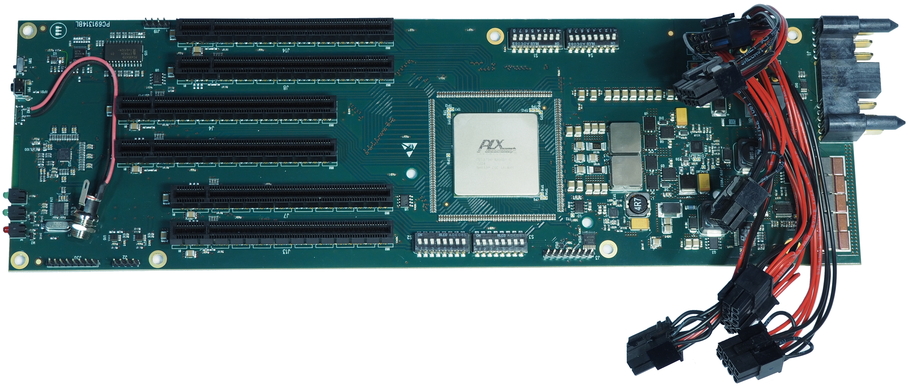}}\hspace*{-8mm}
  \raisebox{56mm}{\includegraphics[height=22mm,rotate=220]{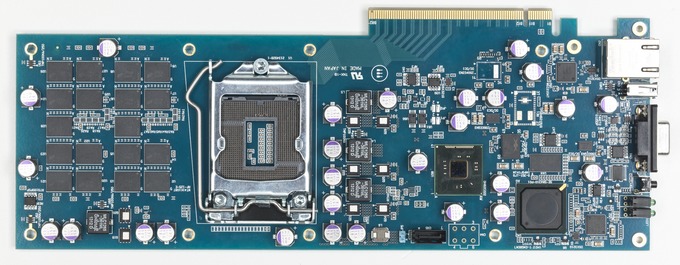}}\hspace*{-7mm}
  \raisebox{5mm}{\includegraphics[height=20mm,rotate=40]{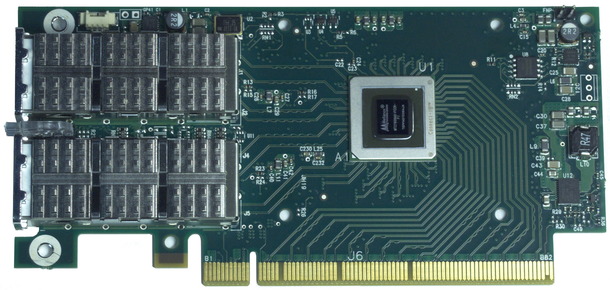}}
  \vspace*{-5mm}
  \caption{Some components of a QPACE 2 node (not to scale): Midplane
    (left), CPU card (middle), and Connect-IB card (right).  The power
    cables on the midplane are needed for the KNCs.  See text for 
    details.}
  \label{fig:mars}
\end{figure}

\begin{figure}
  \centering
  \includegraphics[height=50mm,valign=c]{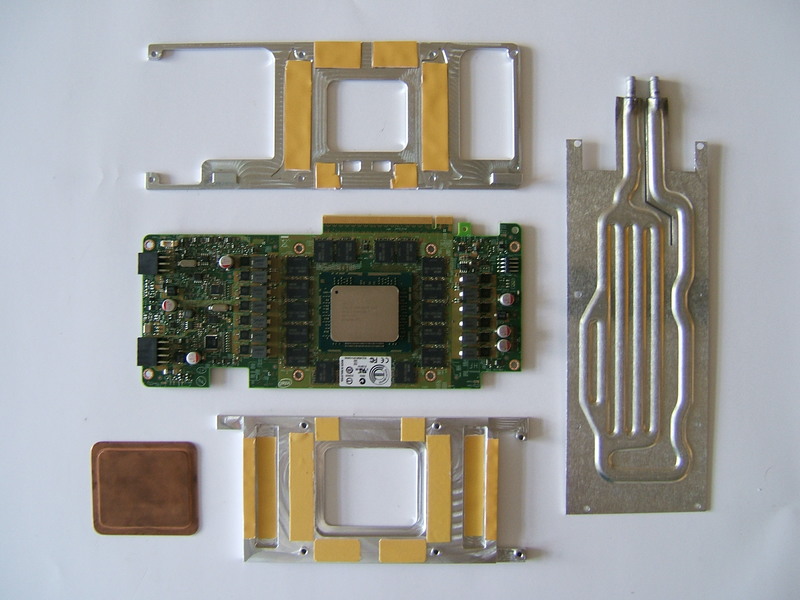}\hfill
  \includegraphics[height=40mm,valign=c]{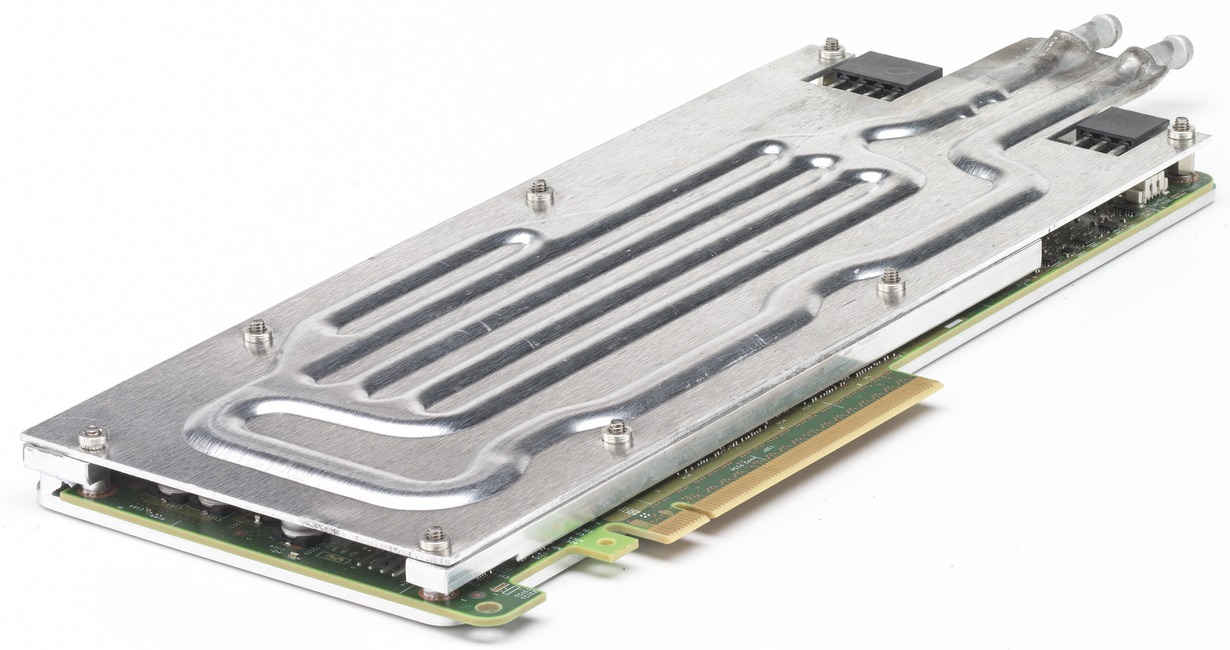}
  \caption{QPACE 2 cooling concept, here for the KNC card.  Left: The
    hot components of the card (middle) are thermally coupled to the
    roll-bond plate (right) via interposers (top and bottom).  The
    copper plate is inserted in the hole of the interposer and
    interfaces with the KNC chip through thermal grease.  The yellow
    pieces are thermal interface material.  Right: Fully assembled
    water-cooled KNC card.  The top interposer is coupled to the
    roll-bond plate with thermally conductive glue.  The bottom
    interposer, which cools memory chips on the bottom of the card, is
    thermally coupled to the top interposer via aluminum surfaces
    (visible in Fig.~\protect\ref{fig:test} right) and screws.}
  \label{fig:knc}
\end{figure}

The components of a node are integrated in what we currently call a
``brick''.\footnote{This name is due to the obvious similarity of our
  mechanical design, see Fig.~\ref{fig:brick}, with a brick used in
  the construction industry, but we should probably change it in the
  future to avoid confusion with the alternative meaning of
  ``non-functioning electronic device''.}  The basis of a brick is a
midplane, see Fig.~\ref{fig:mars} (left).  This midplane accommodates
the PCIe switch, a power connector and a number of power converters,
and six slots for standard PCIe form factor cards, i.e., a CPU card
(Fig.~\ref{fig:mars} middle), a Connect-IB card (Fig.~\ref{fig:mars}
right), and four KNC cards (Fig.~\ref{fig:knc}).
The KNC card (by Intel) and the Connect-IB card (by Mellanox) are
commercial components.  The midplane and the CPU card were designed
for QPACE 2 but can be reused for other projects/products.  The CPU
card also contains 16 GB of DDR3 memory, a platform controller hub
(PCH), and a baseboard management controller (BMC).  It provides two
Ethernet and several other interfaces such as USB, VGA, and SPI.  One
of the Ethernet interfaces is used by the BMC and the other by the CPU
to form two separate Ethernet networks.  The BMC network is used for
low-level control of the nodes.  The CPU network is used for booting
the operating system and for login.  Both networks are also used for
system monitoring.

The cooling concept of QPACE 2 is a novel idea based on roll-bond
technology (see p.~253 and Fig.~8.23 of \cite{beddoes1999principles}
for an explanation of this technology).  The concept is shown in
Fig.~\ref{fig:knc} for a KNC card.  The other cards as well as the
midplane are also cooled in this way.
Water flows through a roll-bond plate (made of aluminum) which is
thermally coupled to the hot components via interposers (made of
aluminum or copper) and thermal grease or thermal interface material.
The advantage of the roll-bond technology is that it is cheap and
widely available and that custom-shaped water channels are very easy
to implement.  In addition, the risk of leakages is minimized.  With
other technologies it is much harder or more expensive to create
channels of arbitrary shape and to make sure that there are no
leakages.  Our design allows for cooling water temperatures of at
least 50$^\circ$C, which makes free cooling possible year-round.  Thus
no chillers are needed, which in our installation improves the overall
energy efficiency of the machine significantly.

\begin{figure}
  \centering
  \includegraphics[height=45mm]{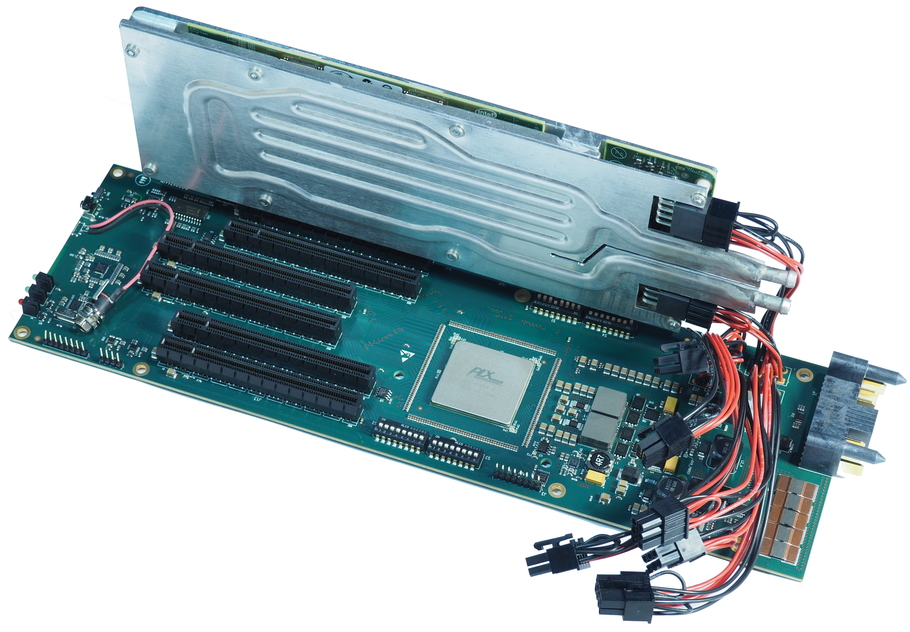}\hfill
  \includegraphics[height=42mm]{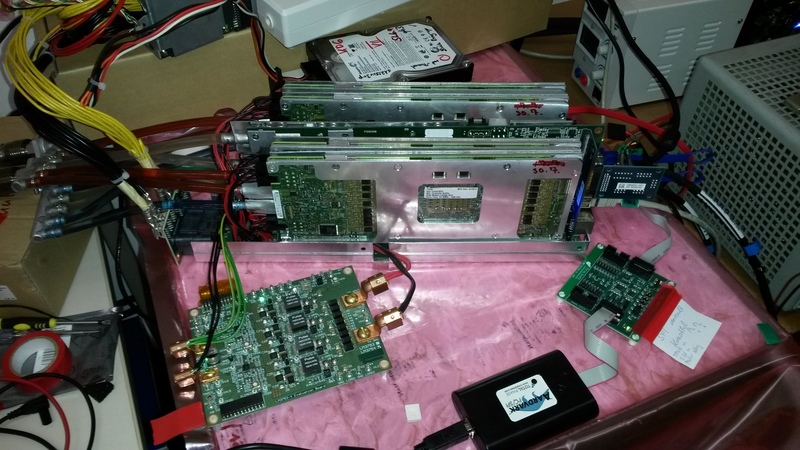}
  \caption{Left: The PCIe cards with their cooling assembly are
    plugged into the midplane.  Here only one KNC card is shown.
    Right: One node (with all six cards liquid-cooled but without
    housing) is undergoing testing.}
  \label{fig:test}
\end{figure}

\begin{figure}
  \centering
  \includegraphics[height=47mm]{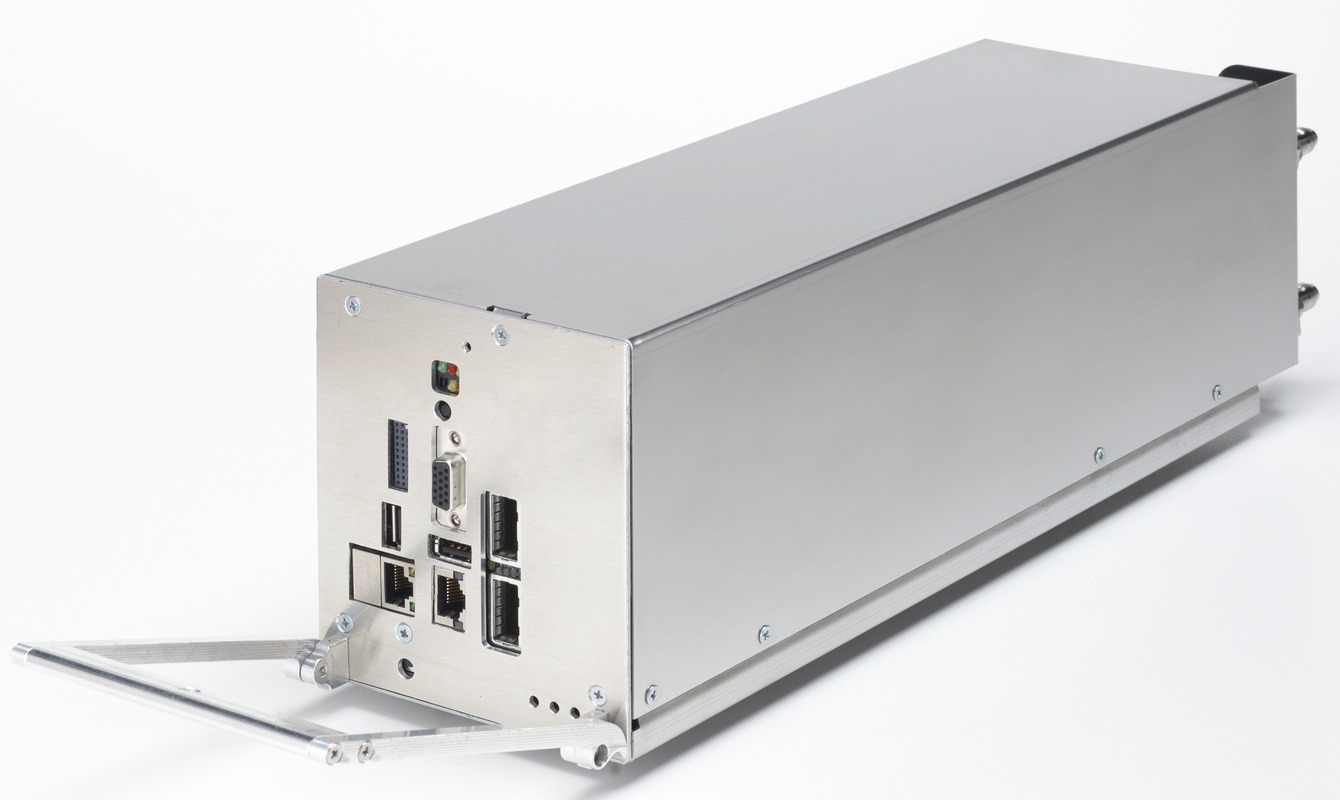}\hfill
  \includegraphics[height=47mm]{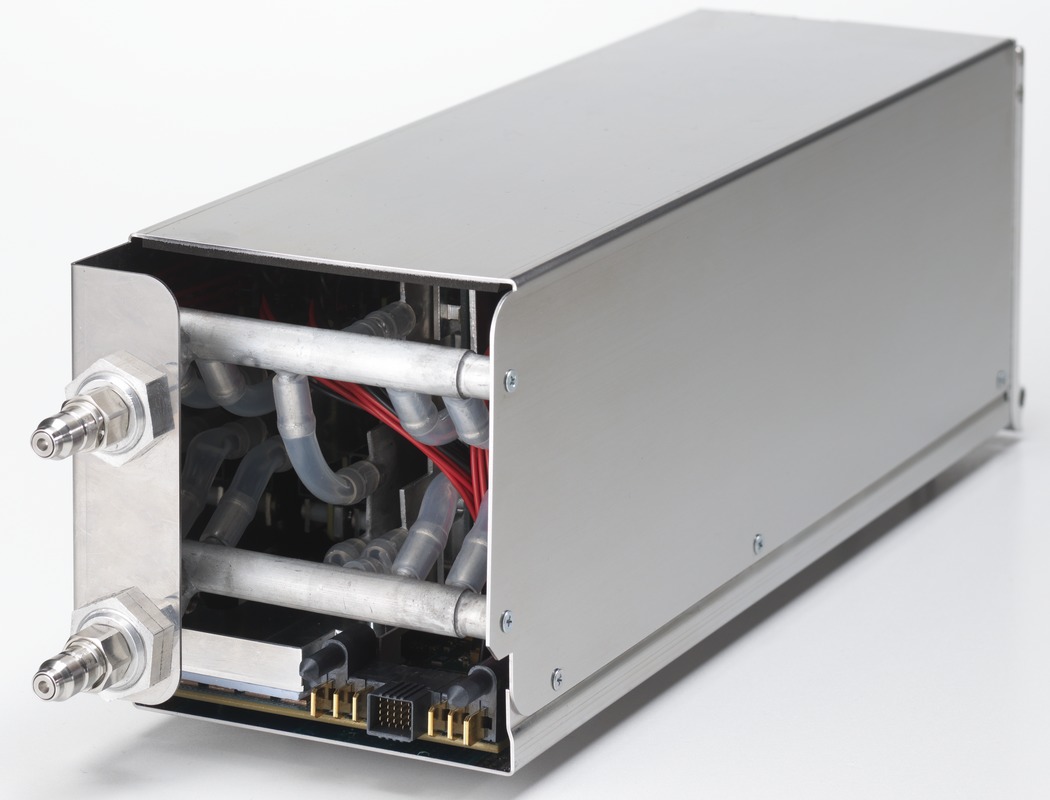}
  \caption{Left: A fully assembled node in its housing.  Several
    interfaces are provided on the front panel: dual FDR Infiniband,
    dual Ethernet, USB, VGA, and debug connector.  Right: Back side of
    the brick with power connector, mini-manifold, and drip-free
    quick-disconnect couplings (male parts) for in- and outlet.}
  \label{fig:brick}
\end{figure}

Once the cooling solutions have been assembled, the six PCIe cards are
plugged into the midplane, as shown in Fig.~\ref{fig:test}.
The six roll-bond plates are connected to a mini-manifold with plastic
tubes.  The entire assembly is then put into a housing, see
Fig.~\ref{fig:brick}. 
The mini-manifold is visible in Fig.~\ref{fig:brick} (right).  It is
connected to the cooling-water circuit via drip-free quick-disconnect
couplings.  The lever visible in Fig.~\ref{fig:brick} (left) locks the
power connector and the quick disconnects into place in the rack.

We use a standard 19-inch rack of height 42U.  The height of a brick
is 3U, and we can put $4+4$ bricks in 3U (4 from the front of the
rack, 4 from the back).  This translates to 64 nodes, i.e., 256 KNCs,
in 24U.  The remaining space in the rack is used for the cooling-water
distribution, power distribution, 3 Ethernet and 4 Infiniband
switches, a management/login server, and cable management.  The peak
performance of a rack is about 310 TFlop/s (double precision).

\begin{figure}
  \centering
  \includegraphics[height=40mm]{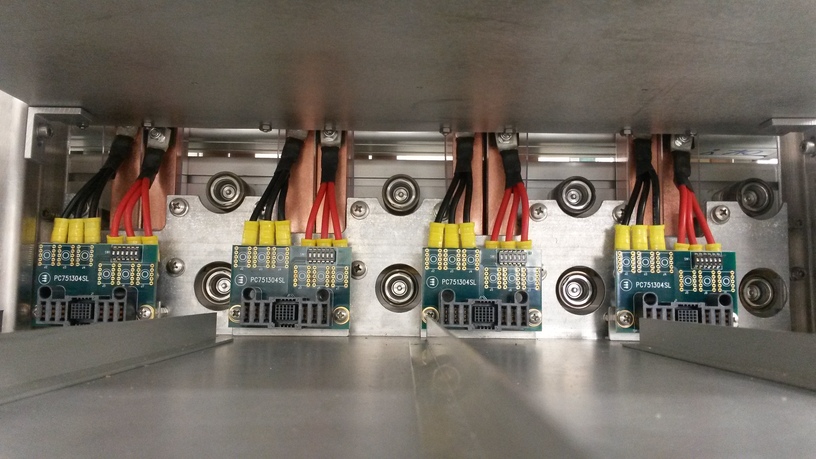}\hfill
  \includegraphics[height=40mm]{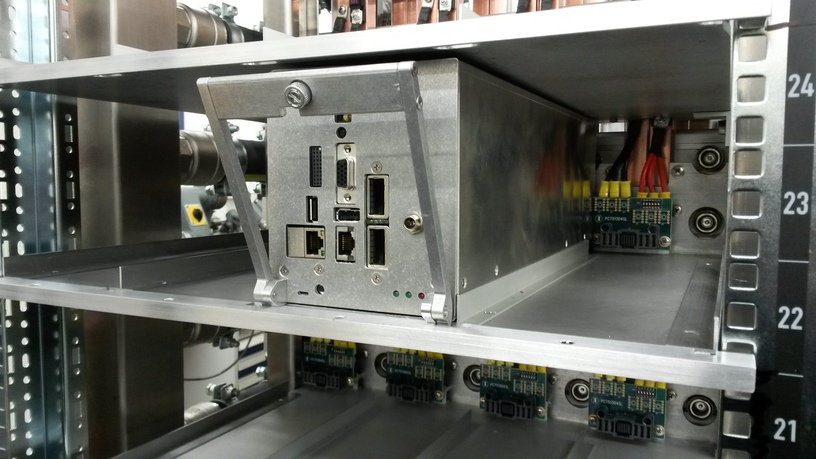}
  \caption{Left: Power bars with power backplanes attached via cables.
    Also visible are the quick-disconnect couplings (female parts) for
    in- and outlet.  Right: A brick (of height 3U) is being inserted
    into its slot.}
  \label{fig:rack}
\end{figure}

\begin{figure}
  \centering
  \includegraphics[height=43mm]{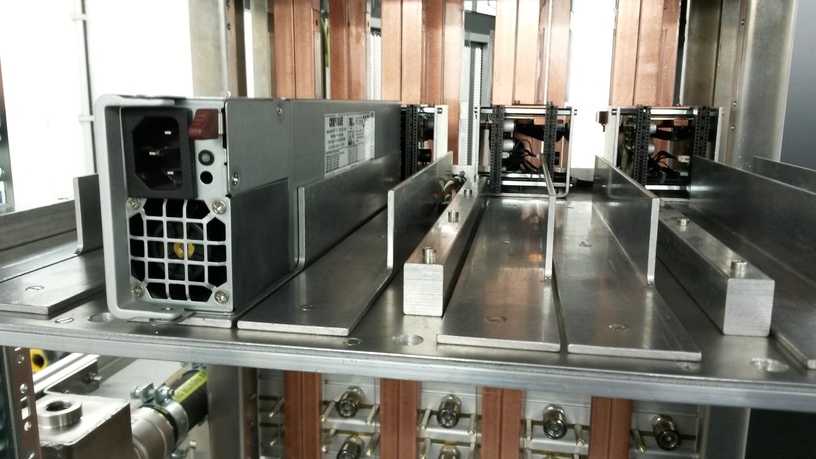}\hfill
  \includegraphics[height=43mm]{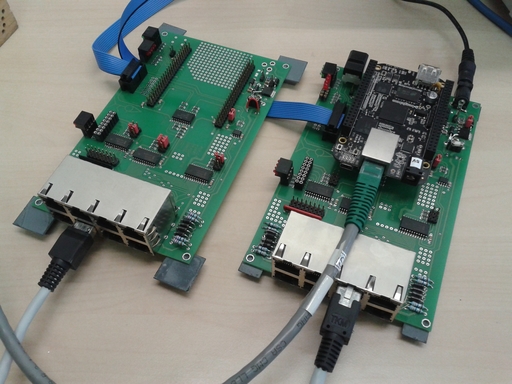}
  \caption{Left: The PSUs are connected to the power bars via power
    distribution boards (PDB).  Each PDB holds two PSUs.  Here, only a
    single PSU is shown.  Right: The master PSU control board (right)
    accommodates a BeagleBone Black single-board computer.  The other
    PSU control boards (e.g., the one on the left) run in slave mode.}
  \label{fig:heinzelmann}
\end{figure}

12V power is delivered to the nodes via massive power bars made of
copper and small power backplanes, see Fig.~\ref{fig:rack}.  The
latter also contain a DIP switch that is used to assign a unique
location ID to each brick.  Based on this location ID we then define
IP addresses.  The power bars are fed by industry-standard platinum
power supply units (PSUs) that convert 230V AC to 12V DC at about 95\%
efficiency, see Fig.~\ref{fig:heinzelmann} (left).  Each PSU can
deliver up to 2~kW.  We subdivide the rack power distribution into 8
domains, with each domain powered by 6 PSUs serving 8 bricks in
parallel.  Assuming a typical power consumption of 1~kW per node, we
are just on the border between $5+1$ and $4+2$ redundancy.  A PSU
control board designed in Regensburg monitors and controls the PSUs
via PMBus, using a BeagleBone Black single-board computer that plugs
into the master control board, see Fig.~\ref{fig:heinzelmann} (right).
One PSU control board services 16 PSUs, hence we have 3 control boards
for 48 PSUs.  The PSUs can deliver a maximum power of 96~kW per rack.
First results suggest that synthetic benchmarks will consume up to
85~kW, whereas high-performance QCD code will run at about 60~kW.

\subsection{System software}

The BMC on the CPU card is running an embedded Linux version and boots
from flash memory as soon as power is turned on.  It supports the
typical functionalities of a board management controller, e.g., it can
hold the other devices in the brick in reset or release them, it can
monitor voltages as well as current and temperature sensors and act
accordingly (e.g., by performing an emergency shutdown), or it can
access the registers in the PCIe switch via an I$^2$C bus.  Note that
our nodes are diskless.  Once the CPU is released from reset it
PXE-boots a minimal Linux image over the Ethernet network.  The full
Linux operating system (currently CentOS 7.0) uses an NFS-mounted root
file system.  The KNCs are booted and controlled by the CPU using
Intel's KNC software stack MPSS.  The KNCs support the Lustre file
system of our main storage system, which is accessed over Infiniband.
A variety of system monitoring tools are running either on the BMC (as
described above) or on the CPU, which, for example, regularly checks
the temperatures of all major devices (KNCs, CPU, Infiniband HCA, PCIe
switch) as well as various error counters (ECC, PCIe, Infiniband).  A
front-end server is used to NFS-export the operating system, to
communicate with the nodes and monitor them, to log in to the machine,
and to control the batch queues.

\newpage

\subsection{KNC architecture and its implications for the programmer}
\label{sec:micro}

Let us first describe some relevant details of the KNC architecture,
focusing on the 7120X version used in QPACE 2.  The processor chip
contains 61 compute cores running at 1.238 GHz.  The cores are
connected by a bidirectional ring bus, which also interfaces to the
GDDR5 memory controllers (for a total of 16 GB of on-board memory) and
the PCIe I/O logic (16 lanes PCIe Gen2).  Each core has a
512-bit wide SIMD unit on which it can perform fused-multiply-add
instructions in either single or double precision.  This translates to
a peak performance of 1.2 TFlop/s per KNC in double precision.  There
are four threads per core that have their own register set but shared
execution units and caches.  Each core has a private L2 cache of 512
kB.  The L2 caches are kept coherent with a distributed tag directory,
i.e., if core A needs data that are not in its L2 but in the L2 of
core B, these data are automatically fetched from core B via the ring
bus, rather than from memory.  This is faster than memory access, but
not by much in practice.

While the KNC architecture is quite promising, the actual hardware
implementation has some performance-relevant pitfalls that one should
be aware of.  For example, instructions from a given thread can only
be issued every other cycle.  Therefore we need at least two threads
to issue instructions in each cycle.  Also, there is no L1 hardware
prefetcher.  Furthermore, the cores execute instructions in order,
which implies that cache misses always stall execution of a thread for
$10\sim100$ cycles.  To hide these stalls we also need to use multiple
threads.

After initial theoretical performance analysis, we have performed
microbenchmarks and experimented heavily while implementing and
optimizing our software.  We believe to have arrived at a solid
understanding of how to write high-performance code for the KNC.  The
purpose of this subsection is to describe some of the insights we have
gained, starting from a single core and going to the full multi-node
system.

For some parts of the application, such as the stencil computation for
the Wilson Dirac operator, the small size of the L1 cache (32 kB) can
become a limiting factor.  There are two reasons for this.  First,
while the L1 size has been kept constant, the size of the vector
registers has been doubled compared to AVX (which in turn is double
that of SSE).  As a consequence the size of the working set is
typically larger, which leads to more L1 capacity cache misses, i.e.,
data are more frequently evicted from L1 before they can be reused.
Second, to fully use a core we need to use two threads, and to hide
stalls due to cache misses we would often like to use three or four.
Thus the L1 size per thread is less than 32 kB.  The 8-way
associativity of the L1 may further reduce the available space per
thread since multiple threads compete for the 8 ways (unless they work
on the same data).  The lack of an L1 hardware prefetcher also means
that software prefetching can be crucial.  With well-placed L1
software prefetches\footnote{The icc compiler does well in this
  respect for simple code parts like loops, but manual prefetches are
  beneficial for more complex parts like the application of the Dirac
  operator.}  we often get good results with two threads.  Without
optimized L1 prefetching typically three or four threads yield the
best performance, at the cost of incurring the negative effects
described above.

The L2 cache is coherent but not shared between cores.  This has
several consequences.  In a typical stencil computation each core
would work on a distinct block of the lattice that needs to be held in
cache.  In addition, also neighboring sites (typically denoted as
halos or ghost shells) need to be accessed, and thus be loaded into
the cache.  Without a shared cache this implies duplication at the L2
level: data for each lattice site may be present in the L2 caches of
several cores.  Similarly, the application code will by duplicated
$\Nc\sim60$ times, since each core needs a copy of it.  Finally,
barriers are very expensive, up to 10,000 cycles for optimized
implementations (see, e.g., an implementation by Intel as part of the
QPhiX library \cite{QPhiX}), while many standard implementations (such
as in OpenMP and pthreads at the time of this writing) need well above
10,000 cycles.

As described above, a substantial fraction of each core's L2 is
``wasted'' for duplication of code and halos, thus reducing the
fraction available for local data.  Therefore we typically need a
considerable amount of main-memory access.  Stream benchmarks yield on
the order of 170~GB/s for the sum of read and write bandwidth (i.e.,
less than 50\% of the nominal 352 GB/s).  This bandwidth can be
sustained for simple cases like streaming access thanks to the L2
hardware prefetcher.  For more complex cases software prefetching may
become necessary.  The general memory access pattern is of importance
as well.  To obtain a bandwidth close to 170 GB/s, all cores must read
and/or write simultaneously and continuously, i.e., if each core reads
and/or writes in bursts (with breaks in between) the resulting overall
bandwidth is much lower.  Furthermore, if each core reads streams from
more than about $5\sim10$ memory locations at a time, the overall
bandwidth will not reach 170 GB/s, not even with manual L2
prefetching.  This may be alleviated by modifying the data layout so
that data needed at the same time are placed contiguously, e.g., by
packing the gauge links for all four directions in one array instead
of keeping a separate array for each direction.

For moving data over the network we typically rely on MPI, which in
our case can use Intel's SCIF library (for intra-node communication)
and/or the Infiniband HCA (for intra- and inter-node communication).
SCIF shows good results in microbenchmarks: RDMA between two KNCs
sustains a bidirectional bandwidth of about 6.5 GB/s for payloads of
64 kB and above, while PIO (programmed IO) sustains about 5 GB/s for
payloads of 16 kB and above.  These numbers are to be compared to the
nominal value of 7.2 GB/s, which takes into account the overhead for
the 256-Byte PCIe packages supported by the KNC \cite{PLX}.  The
zero-size latencies are about 5.5~$\upmu$s for RDMA and about
2~$\upmu$s for PIO.  However, in practice we experience issues: when
using SCIF RDMA calls, the kernel seems to heavily utilize $5\sim10$
cores to do the data movement.  These cores then become unusable for
computation, since the application threads on these cores would slow
down the whole machine.\footnote{One solution would be to leave these
  cores free, giving up 15\% of the total performance, but even that
  is not enough: the kernel may occasionally schedule SCIF-related
  threads on one of the ``compute'' cores, again causing slowdown. In
  a big system this will of course happen very frequently.  The
  situation could be improved by pinning these SCIF-related kernel
  threads to specific cores.  This would be an interesting option for
  future work.}  Intel MPI
% \TW{version 4.1, 5.0 has issues with dual-port IB cards}
via the HCA, which utilizes the DMA engine of the HCA rather than that
of the KNC, does not show these issues.  Another point is that the
latency induced by the Intel MPI software stack is quite high.  In
microbenchmarks MVAPICH2-MIC
% \TW{version 2.0}
seems to perform a bit better than Intel MPI (see also
\cite{Potluri:2013:MPC:2503210.2503288}), but at the time of this
writing MVAPICH2-MIC always shows the SCIF issue explained above, even
in cases where only the HCA should be used.  In some
performance-relevant parts we were successful in improving the
performance by using Infiniband directly (i.e., without MPI) via the
IB verbs library \cite{IBverbs}, at the cost of significantly
increased complexity for the programmer.

\subsection{Application software}
\label{sec:app}

We mainly use Chroma \cite{Edwards:2004sx}.  For the
performance-relevant parts such as the DD-based solver described in
Sec.~\ref{sec:DD} we wrote separate libraries that can be linked
during Chroma compilation.  The Wuppertal adaptive algebraic multigrid
code \cite{Frommer:2013fsa,Frommer:2013kla} has been ported to the
KNC, including vectorization of all performance-relevant parts.  In
this code a Krylov solver is preconditioned with a V-cycle and a
smoother. For the latter we use our DD preconditioner.  Other parts of
our code suite, such as HMC, hadron spectra, and hadron distribution
amplitudes are also being adapted to the KNC.  Furthermore, we made
some improvements to QDP++ \cite{Edwards:2004sx}, such as memory
management, optimization of existing threading, and some new
threading.  As for the actual implementation, the code is written in
C++, the performance-relevant parts are vectorized using icc
intrinsics or auto-vectorization by the compiler, threading uses
OpenMP, and the multi-node implementation is currently based on Intel
MPI or IB verbs as described in Sec.~\ref{sec:micro}.

\section{Domain decomposition on the Knights Corner processor}
\label{sec:DD}

Most of the content of this section is a summary of work done in
collaboration with the Intel Parallel Computing Labs in Santa Clara
(USA) and Bangalore (India) and with JLAB (USA), the details of which
are described in \cite{Heybrock:2014iga}.  This work was motivated by
the first implementation of a conjugate-gradient-based solver on the
KNC \cite{Joo:2013a} (see also
\cite{Li:2014kxa,Jang:2014mxa,Kaczmarek:2014mga} for other
implementations).  The strong-scaling behavior of that solver on a
production system (TACC Stampede) turned out to be rather limited, the
performance bottleneck being the bandwidths for memory access and
inter-node communication.  The obvious way to improve upon the
strong-scaling behavior is to switch to another algorithm, based on
domain decomposition, that moves less data between KNC and memory, and
between different KNCs.

In fact, this is a good example of how the implementation of
high-performance application code on a new hardware architecture can
proceed.  First, one identifies the most suitable algorithm for the
given hardware architecture.  Second, one identifies the most suitable
data layout for the given algorithm and hardware architecture.  Only
then should one proceed to optimize the code (including vectorization,
threading, parallelization, etc.).

\subsection{Algorithm and implementation overview}

Let us briefly describe some details of our algorithm.  For the Dirac
operator we use Wilson-Clover.  Our outer solver is flexible GMRES
with deflated restarts \cite{Frommer:2012zm}.  As a preconditioner we
use the multiplicative Schwarz method (a.k.a.\ Schwarz alternating
procedure), which was introduced in \cite{Schwarz:1870} and adapted to
Lattice QCD in \cite{Luscher:2003qa}.  The main idea of the Schwarz
method is to subdivide the lattice into domains.  After a reordering
of indices, the matrix to be inverted then consists of a
block-diagonal part (corresponding to the interactions within the
domains) and a remainder (corresponding to the interactions between
domains).  Based on this splitting, a block-Jacobi iteration can be
performed to approximate the inverse.  If the domains are small enough
to fit in cache, the ensuing inversion on the domains (i.e., the
block-diagonal parts) can be done without off-chip data
movement.\footnote{For the inversion on the domains we use the
  minimal-residual (MR) algorithm with even-odd preconditioning.}
Computation of the remainder does require communication but occurs
less frequently.  As a result, the overall algorithm needs less
communication bandwidth, is more latency tolerant, and gives more
cache reuse than the conjugate-gradient algorithm.

\subsection{Data layout and vectorization}

For optimal performance it is necessary (i) to fully utilize the SIMD
units, (ii) to choose the data layout such that the cache lines loaded
into the KNC do not contain unnecessary data, and (iii) to avoid
instruction overheads due to SIMD vector permutations.  Typically one cannot
satisfy all criteria simultaneously and tries to find a compromise
that maximizes the overall performance.  In our case, the best
solution is a structure-of-array (SOA) format in which all 24
floating-point components of a spinor are stored in 24 separate
registers and cache lines.  This leads to ``site fusing'', i.e., one
512-bit register contains data from 16 sites (in single precision).
Since we have one small domain per core (see Sec.~\ref{sec:cache}) we
have to site-fuse in more than one dimension.  For example, for an
$8\times4^3$ domain we site-fuse the $x$- and $y$-dimensions, with
data from $4\times4$ sites in each register.\footnote{Our notation is
  always $L_x\times L_y\times L_z\times L_t$, i.e., $8\times4^3$ means
  $L_x=8$ and $L_y=L_z=L_t=4$.}

In this scheme, the computation of hopping terms within domains can be
done straightforwardly in the $z$- and $t$-directions for complete
registers, while in the site-fused directions some permute and mask
instructions are needed that lead to an underutilization of the SIMD
units of 12.5\% in the $x$- and 25\% in the $y$-direction.  For the
computation of hopping terms between domains, again the $z$- and
$t$-directions are straightforward, while in the site-fused directions
the cache lines that need to be loaded to access the neighbor's
boundary data contain many data that are not needed.  We avoid this
overhead by additionally storing the boundary data in an
array-of-structure (AOS) format \cite{Heybrock:2014iga}.

\subsection{Cache management and prefetching}
\label{sec:cache}

Since the L2 is not shared we assign each domain to a single core.
The cache size of 512~kB per core restricts the domain size to
$8\times4^3$ (in single precision).  We actually implemented domains
of size $4^4$ and $8\times4^3$.  Since the KNC can do up- and
down-conversion, we store some domain data, i.e., the gauge links and
clover matrices, in half precision.  This reduces the working set
(from 456~kB to 312 kB per $8\times4^3$ domain) and the
memory-bandwidth requirements.  To ensure stability, we keep the
spinors in single precision.  We observed that this optimization had
no noticeable impact on the iteration count of the outer solver.

As discussed in Sec.~\ref{sec:micro}, the KNC has no L1 hardware
prefetcher, and an L2 hardware prefetcher only for streaming access.
Because of the irregular code structure, compiler-generated software
prefetches are often not good enough.  It is therefore essential to
manually insert L1 and L2 prefetches using intrinsics.  The
fine-tuning of these prefetches turned out to be rather
time-consuming.

\subsection{Intra-core threading}

As mentioned in Sec.~\ref{sec:micro}, at least two threads per core
are necessary to fully utilize the floating-point unit.  In our
implementation we assign threads to alternating time slices within a
domain.  There are pros and cons for using two or four threads per
core.  Two threads suffer from more stalls since L1 or L2 misses
cannot be hidden by other threads.  Four threads suffer from more L1
conflict misses because our working set exhausts the L1 size.  We
measured the performance for both cases and found no significant
differences.

\subsection{Inter-core parallelization}
\label{sec:intercore}

When parallelizing over the cores of the KNC, synchronization between
cores is only necessary after an MR block solve, which implies that
the cost of a barrier (which can be quite large on the KNC, see
Sec.~\ref{sec:micro}) has little impact on the overall performance.

A simple but very performance-relevant issue is load balancing.
Standard lattice sizes are typically powers of 2 and thus do not map
naturally to 60 or 61 cores.  As a result, some of the cores are
unused at least part of the time (``load imbalance'').  A simple
solution would be to include factors of 3 and 5 in the lattice size,
but this is only an option for new simulations where we can select the
lattice size.  Another option is to
increase the local volume (per KNC) so that the number $\Nd$ of
domains assigned to a single KNC\footnote{Note that the number of
  domains that can be processed independently is given by $\nd=\Nd$/2.
  The factor of 2 is due to the checkerboarding in the multiplicative
  Schwarz method.}  is much larger than the number $\Nc$ of cores per
KNC.  The average load is then given by the function $x/\ceil(x)$ with
$x=\Nd/2\Nc$ \cite{Heybrock:2014iga}, which has a saw-tooth-like
behavior with the minima approaching 1 from below as $\Nd$ increases.
However, increasing the local volume is not an option in the
strong-scaling region for which we are trying to optimize.  The third
option is a non-uniform distribution of the lattice over multiple KNCs
(assuming a multi-node implementation as described below).  Let us
give a simple example to illustrate the idea, using a lattice size of
$64^3\times128$ and a domain size of $8\times4^3$.  If the lattice is
uniformly distributed over, say, 1024 KNCs in a
$4\times4\times8\times8$ layout, we have $\Nd=64$ and thus a load of
$53\%$.\footnote{Here, we use $\Nc=60$ since Intel configures the
  Linux kernel to run on the last core.}  In that case each KNC
contains $16$ sites in the $t$-direction.  Alternatively, we could
split the $t$-direction non-uniformly as $128=4\cdot28+16$ so that we
need only 5 (instead of 8) KNCs for the $t$-direction.  This reduces
the number of KNCs from 1024 to 640 and increases the average load to
$(4\cdot56+32)/(5\cdot60)=85\%$.  The time to solution will not
decrease (and may in fact increase slightly because the boundaries to
be communicated will be larger), but the cost to solution will
decrease by almost 40\%.

\subsection{Multi-node implementation}

For the parallelization over many KNCs using MPI we have several
options.  For example, each thread could issue its own MPI calls.
However, this option has two disadvantages.  First, the overhead of
many simultaneous MPI calls is high in typical MPI implementations.
Second, the packets generated by each thread are too small to
efficiently utilize the network.  Therefore, a better option is to
combine the surface data of all domains and to communicate them using
only a single thread.  This option also has a disadvantage, i.e., the
need for explicit on-chip synchronization, but in practice it performs
better than the other option.

As in any parallelization effort, it is essential to overlap
computation and communication in order to hide the communication
latencies.  The standard method is to divide the local volume into
interior and surface and to work on the interior while waiting for the
surface communication.  For our domain-decomposition approach this
does not work for typical parameters since most domains would be on
the surface, i.e., we have only a small interior, or none at all.
\begin{figure}
  \centering
  \includegraphics[height=40mm,valign=c]{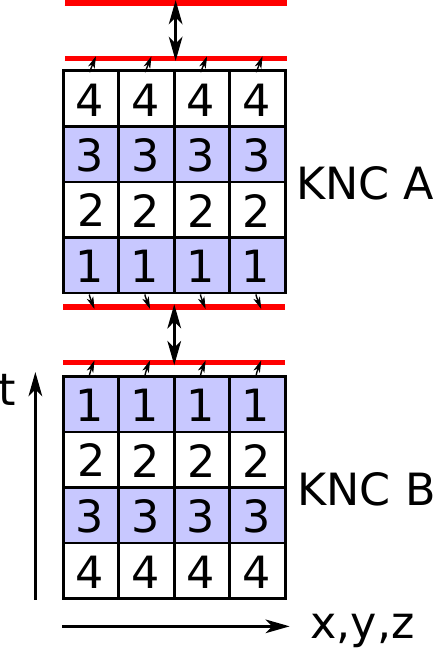}\hfill
  \includegraphics[height=40mm,valign=c]{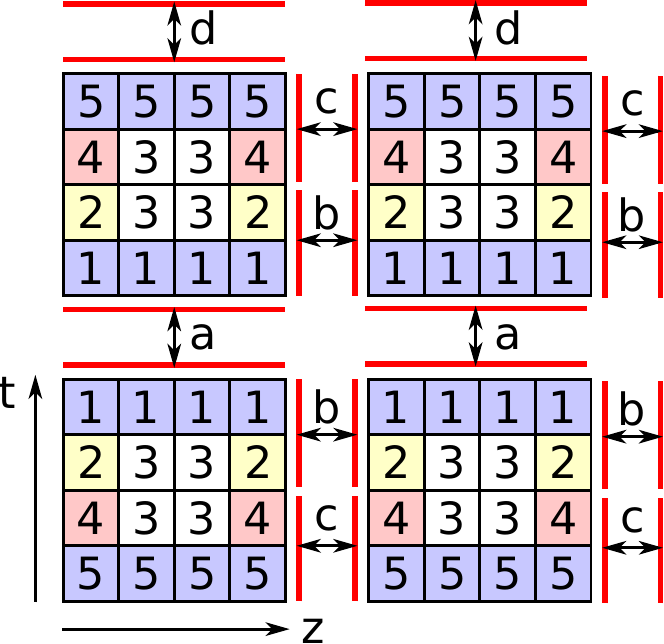}\hfill
  \includegraphics[height=25mm,valign=c]{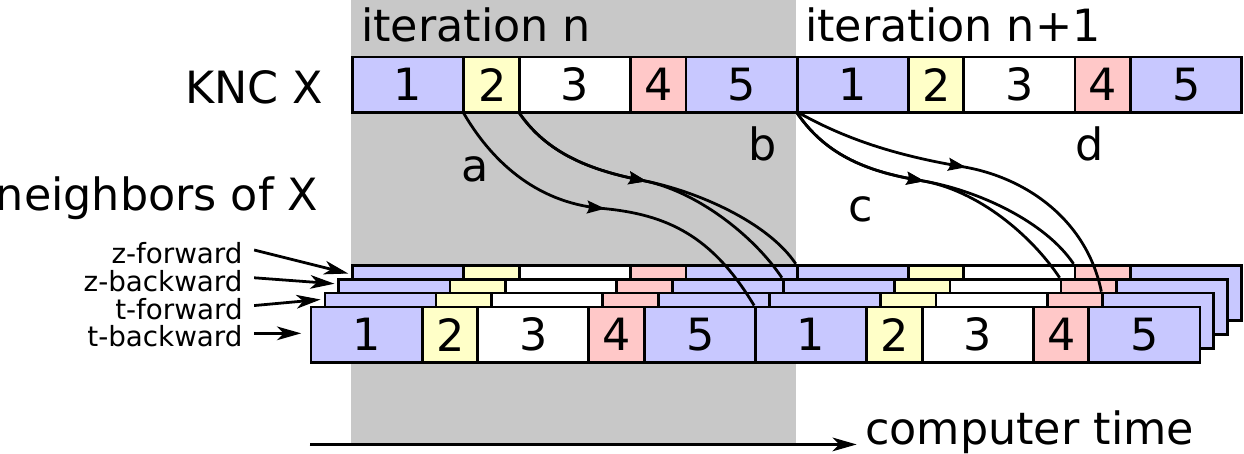}
  \caption{Illustration of our method to hide communication latencies.
    Each little box represents a domain.  Left: Splitting only in the
    $t$-direction.  Middle: Splitting in $t$- and $z$-direction.
    Right: Linear representation of the scheme in the middle.  See
    text for details.}
  \label{fig:multinode}
\end{figure}
Instead, we devised a new method, see Fig.~\ref{fig:multinode}.  In
the left part we first show a splitting in a single direction, here
$t$.  Computations are done on each time-slice in the order indicated
by the numbers.  Boundary data can be sent after (1) has finished.
These data are only needed for (1) in the next iteration.  Thus, the
communication of (1) overlaps with the computation of (2)--(4).
However, this only works for $t$, but not for the other directions
because their boundaries can only be sent after (4) has finished.  A
better scheme is shown in the middle and right parts.  Here, we split
in the $t$- and $z$-directions (in practice we split in all
directions, and the same principle applies).  Again, the order of the
computations is indicated by the numbers, while the order of the
communications is indicated by letters.  As before, (a) occurs after
(1), but in the $z$-direction we now send boundary data when the
computation on half of the boundary has finished, i.e., (b) occurs
after (2).  The other half of the boundary is sent at the end, i.e.,
(c) occurs after (5).  The first half is needed at the start of the
next iteration, which means that (b) overlaps with (3)--(5).  The
second half is needed before (4) of the next iteration so that (c)
then overlaps with (1)--(3).

\subsection{Results}

In this subsection we discuss the performance of our code for three
cases: single core, single KNC, and many KNCs.  Since we are
interested in realistic lattice sizes, for which the lattice is
distributed over many KNCs, the relevant parts of the code are MR
iteration and DD preconditioner for the single-core and single-KNC
cases, and full solver for the many-KNC case.

We first present some theoretical single-core performance estimates
for the Wilson-Clover operator, all in single precision.  We assume
that our domains fit in cache so that we are not
memory-bandwidth bound.  Of the 1848 flops/site in Wilson-Clover, 64\%
are fused multiply-adds, which gives an upper performance limit of
82\%.  Instruction overheads due to masking, shuffles and permutes as
well as imperfect instruction pairing reduce the theoretical limit to
56\% or 22 Gflop/s/core, see \cite{Heybrock:2014iga} for details.  Our
measured performance for the MR iteration is about 13.3 Gflop/s/core.
According to VTune (Intel's performance-analysis tool) this
performance loss is mainly due to stalls because of outstanding L1
prefetches.  Since the optimal number of MR iterations in terms of
time-to-solution is only $4\sim5$, the Schwarz method is not completely
dominated by MR, i.e., other parts contribute noticeably, and the
overall performance of the Schwarz method drops to 9.5 Gflops/s/core.
Table~\ref{tab:singlecore} shows the impact of our optimization
efforts on the performance numbers.
\begin{table}
  \centering
    \begin{tabular}{l|cc|cc}
      & \multicolumn{2}{|c|}{MR iteration} & \multicolumn{2}{|c}{DD method} \\
      & single & half & single & half \\
      \hline
      no software prefetching & $6.1$  & $8.9$  & $4.6$ & $6.6$ \\
      L1 prefetches           & $10.4$ & $13.3$ & $6.5$ & $8.7$ \\
      L1+L2 prefetches        & $10.2$ & $13.3$ & $7.1$ & $9.5$ \\
      \hline\noalign{\smallskip}
    \end{tabular}
    \caption{Single-core performance in Gflop/s (single precision)
      for the 7120X. See text for details.}
  \label{tab:singlecore}
\end{table}

\begin{figure}
  \centering
  \includegraphics[height=60mm]{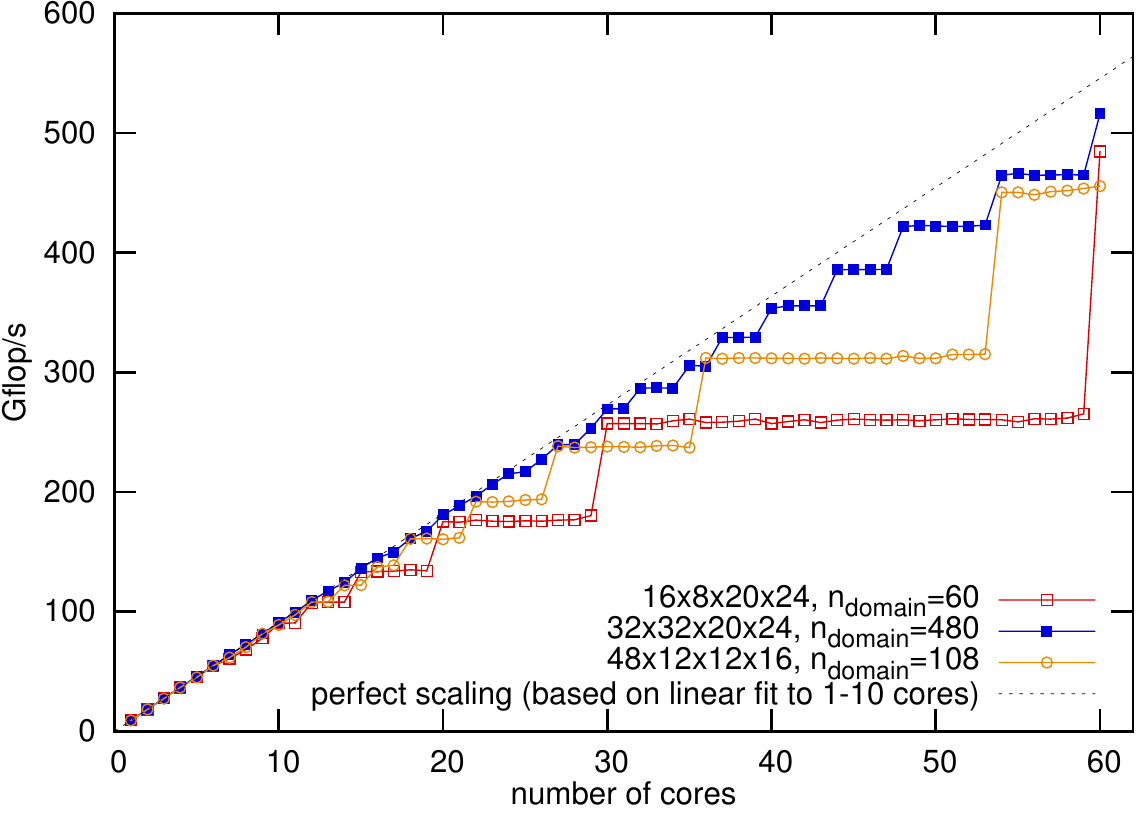}
  \caption{On-chip strong-scaling of the DD preconditioner (with 5 MR
    iterations and 16 Schwarz iterations). The first two volumes are
    chosen such that $\nd$ is divisible by 60, i.e., on 60 cores we
    have 100\% load. The other volume corresponds to distributing
    a $48^3\times64$ lattice over 64 KNCs, with a load of 90\% on 60
    cores.}
  \label{fig:manycore}
\end{figure}

Let us now discuss the many-core performance of the DD preconditioner
when parallelizing over 60 cores of the KNC, see
Fig.~\ref{fig:manycore}.  We clearly observe the load-balancing issue
discussed in Sec.~\ref{sec:intercore}.  As long as the local volume
and thus the number of domains is large enough, the effect is small,
but once $\Nd$ gets close to $\Nc$ it becomes serious.  Apart from
this effect the scaling is almost linear, because (i) during the MR
inversion the cores run independently from their respective caches,
i.e., memory bandwidth does not become a bottleneck, and (ii) the
other (i.e., non-MR) parts of the DD method, which do require
synchronization, are sub-dominant.

Since the full QPACE 2 machine is not yet available, we performed
multi-node benchmarks of the complete solver on up to 1024 nodes of
the TACC Stampede cluster, which uses the 7110P version of the KNC.
Due to the limited space we do not give a detailed discussion here but
refer to \cite{Heybrock:2014iga} for details.  As an example, we
briefly discuss the results for a $64^3\times128$ lattice.  In the
strong-scaling case relevant for HMC, the non-DD solver of
\cite{Joo:2013a} levels off at about 128 KNCs, while our DD solver
still scales roughly linearly up to the maximum partition size of 1024
KNCs, with a time-to-solution gain of about a factor of 5.  In the
weak-scaling (or minimum-cost) case relevant for data analysis, our DD
solver outperforms the non-DD solver by about a factor of $2\sim3$.

\section{Summary and outlook}
\label{sec:concl}

We have presented the architecture of QPACE 2, a massively parallel
and very densely packed system of KNC processors coupled by a
two-layer network based on PCIe and FDR Infiniband.  Two complete
bricks have been running synthetic benchmarks as well as lattice QCD
simulations continuously for over a month without errors.  The full
machine is currently (February 2015) being assembled and tested in
Regensburg and will enter production mode soon.  High-performance
solver code based on domain decomposition as well as application code
is already available, see Sec.~\ref{sec:app}.  The plan is to make
most of our codes publicly available in the near future.

The design of QPACE 2 allows for an upgrade to the next version of the
Xeon Phi processor, code-named Knights Landing (KNL).  Compared to the
KNC, this processor will have three times the floating-point
performance at the same or even less power, better cores (with
out-of-order execution and hardware prefetcher), high-bandwidth memory
integrated on the package, a PCIe Gen3 interface, and in some versions
an integrated network controller (Omni-Path).  A follow-up project,
QPACE 3, will use the KNL processor and have a target peak performance
of about 4 Pflop/s in double precision.  The plan is to install this
system at the Jülich Supercomputing Center in 2016.

%\section*{Acknowledgments}

We thank the German Research Foundation (DFG) for the funding that
made this project possible, the machine shop and the electronics lab
of the Regensburg University Physics Department for their
contributions to the design, and T-Platforms for their support at an
early stage of the project.

\bibliographystyle{JBJHEP_mod}
\bibliography{lat14_plenary}

\end{document}